# THE NON-LINEAR DYNAMICS OF SOCIOLOGICAL REFLECTIONS

Loet Leydesdorff

**Abstract**

**Actors are embedded in networks of communication: the relations of the actors can be represented as the rows of a matrix, while the column vectors represent their communications. The two systems are structurally coupled in the co-variation: each action can be considered as a communication with reference to the network. Co-variation among systems if repeated over time, may lead to co-evolution. Conditions for stabilization of higher-order systems are specifiable: segmentation, stratification, reflection, differentiation, and self-organization can be distinguished in terms of developmental stages of increasingly complex networks. The sociological theory of communication occupies a central position for the clarification of the possibility of a general theory of communication, since it confronts us with the limits of reflexivity in human understanding and reflexive discourse. The implications for modelling the relations among incommensurable discourses (e.g., paradigms) are elaborated.**

## Introduction

Among other things, post-modern criticism has taught us that social structure should not be considered as a given, but as a construct. The demonstration of degrees of freedom in the interaction does not disprove structure, but one can argue that the uncertainty undermines the possibility of a deterministic prediction of structure. Accordingly, Giddens (1979) suggested to consider social structure as an *operation*: social structure is constructed by the aggregation of actions, and at a next cycle --- perhaps after a phase of sedimentation and codification --- structure is 'enabling and constraining' further action. This 'duality of structure' --- as Giddens called the operation --- remains 'virtual' for the actors involved.

More recently, Luhmann (1984) proposed to conceptualize the social system as the network of communications, which is added to the actors who carry it at the nodes. He argued that this communication system --- in analogy to a neural network --- can become self-organizing under specifiable conditions of *cultural* evolution. However, Luhmann's 'social system' remains as much an analytical construct as Giddens' 'virtual structure'. The specification of the network's assumed operation in terms of self-organization therefore has to be considered a second-order hypothesis. However, the empirical researcher would like to see a specification of the conditions for self-organization in the social system.

Can the hypothesis of the self-organization of communication systems further be specified? Is it possible to elaborate Shannon's (1948) mathematical theory of communication in order to account for the dynamics of networks of reflexive communication (cf. Kaufer and Carley 1993; Krippendorff 1994)? In this study I show the fruitfulness of such an abstract specification by applying it to 'virtual' social systems as introduced by Giddens and Luhmann. Among other things, the deductive analysis will enable me to explain why 'unintended consequences' (Giddens 1984) are pervasive in social relations.

**Actors and networks**

A virtual network of links exists by virtue of being carried by actors at its nodes (cf. Rumelhart *et al.* 1986). The two systems are structurally coupled: each action with reference to an actor can be considered as a communication with reference to the network. While the uncertainty is *relational* in terms of the actors involved, it has a *position* in the network (Burt 1982). In other words, the network is spanned in terms of relations, but it develops a specific architecture in which each action has also a position.

The position of a communication is latent for the actors involved (Lazarsfeld & Henry 1968), but the human actor can reflect on each uncertainty, and provide it with a meaning. Meaning requires reflection on the position of information in a system (MacKay 1969). In general, while the relational uncertainty has a position at the interface between two systems, meaning can only be specified with reference to a system for which the incoming information makes a difference over time as a *third* degree of freedom.[1]

Information can be attributed different meanings by different systems. Some systems (e.g., human beings) are able to reflect on various possible meanings, and to make choices among alternatives with hindsight. This operation requires an additional (i.e., *fourth*) degree of freedom. Thus, meaning is itself reflexive (with reference to the dynamics in the first and the second dimension), and it can be made the subject of reflection if it can again be communicated as an uncertainty.

Self-organization can be defined as the ability of a system to select among communicated meanings with reference to the system's identity. While a reflexive meaning can be provisionally stabilized, a further selection among various possible meanings potentially globalizes the system. Thus, the identity of a self-organizing system can be considered as a distributed and changing regime of representations (Hinton *et al.* 1986). One can observe this system in terms of instantiations (Giddens 1979), by taking a position (Haraway 1988) or by using a geometrical metaphor (Shinn 1987). A system which is able to operate in four dimensions of uncertainty, is no longer expected to exhibit observable stability over time.

**Table I**

Organization of concepts in relation to degrees of freedom in the probability distribution

|  | *first* dimension | *second* dimension | *third* dimension | *fourth* dimension |
|---|---|---|---|---|
| *operation organization* | variation | selection | stabilization | self- |
| *nature* | entropy; disturbance | extension; network | localized trajectory | identity or regime |
| *character of operation* | probabilistic; uncertain | deterministic; structural | reflexive; reconstructiv | globally organized; resilient |
| *appearance* | instantaneous and volatile | spatial; multi-variate | historically contingent | hyper-cycle in space and time |
| *unit of observation* | change in terms of relations | latent positions | stabilities during history | virtual expectations |
| *type of analysis* | descriptive registration | multi-variate analysis | time-series analysis | non-linear dynamics |

**Selection**

As noted, a communication network is a piecemeal construct. However, the number of possible links in a network increases with the square of the number of nodes, so that an evolutionarily emerging network is by definition selective with respect to the range of its possible shapings.[2] By repetitive operation, one expects certain linkages to be intensified more than others for stochastic reasons (cf. Arthur 1988). The emerging architecture can be considered the network's structure.[3]

While relations are aggregative and hierarchical, the constructed network can be decomposed with hindsight in terms of its structural components, e.g. in terms of network densities. The decomposition, however, may follow another logic than the network's aggregative composition, since the aggregation has introduced a grouping variable as a second degree of freedom. The grouping that prevails is latent for the actors involved in constructing the network, since each new action may change the grouping rule over time.

**Stabilization**

The relations of the actors can be represented as row-vectors, and the network as a summation of these vectors, i.e., as a two-dimensional probability distribution or a matrix (S pij). In each cycle, the potentially co-varying relations among the actors add up to communications over the columns.

The time dimension adds a third axis to this two-dimensional representation: matrices at different moments in time add up to a cube (S pijk). If one rotates this cube ninety degrees, one can analyze structure in the time dimension, analogously to structure in the matrices at each moment in time. In other words, a communication system contains two structures if it communicates information --- i.e., is contingent --- in the time dimension. The first structure positions the information in the relations on a second dimension of the probability distribution, and the result of this operation can be reflected on the third dimension. Co-occurrences of co-variations can be analyzed as the system's history.

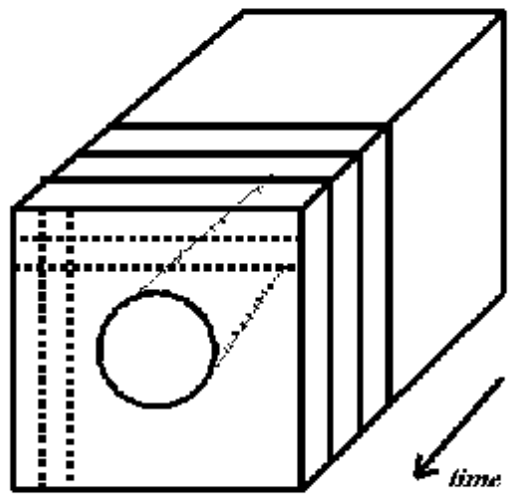

**Figure 1**

*An observable trajectory of a (potentially complex) system in three dimensions*

If the two structures can operate as selections on each other, this may lead to stabilization, e.g., into an observable trajectory (see *Figure One*; cf. Dosi 1982). Thus, stabilization is the result of a second-order selection: which selections are selected for stabilization? Although the two selections are formally equivalent, their orthogonality implies that their operation is substantively independent. I shall argue below that this substantive difference leads to a different semantics in the theoretical understanding.

**Segmentation, stratification and differentiation**

If one represents communicating actors as the row vectors (S pi) of a matrix, communication finds its origin in the co-occurrence of these vectors along the column dimension (*j*). If originally the rows are just stacked upon one another, the result is a segmented communication system. If the communications are ranked in the vertical dimension, one gets a stratified communication system, and grouping of the rows leads to a differentiated communication system. As noted, grouping implies the addition of another dimension to the probability distribution, i.e., a grouping variable *k*. Ranking is the special case that the grouping variable only counts the rank. For example, in a stratified social system a person is allowed to say something if it is his or her turn.

The development pattern is well-known from biology: once the (segmented) *morula* grows so large that there is a need for synchronization among cells no longer directly adjacent, the original symmetry is broken, and order is induced at the system's level. The event of a cell-cleavage is asymmetrically communicated to neighbouring cells, and triggers there a further cleavage. At first, this order is only rank-order, i.e., stratification or, in biological terms, 'polarization' (*gastrula*). The next stage (*blastula*) can be defined as the phase after which undifferentiated cells have ceased to occur.

As long as the windowing of communicating systems on each other remains direct, there is no evolutionary order. The sequencing at the interface induces order. In a stratified system, the communications are ranked at a single centre of reflection, but not yet grouped. Reflection on the distinction between this centre and the periphery using the second (spatial) dimension of the uncertainty induces differentiation. In a self-organizing system, the different meanings can be reflexively communicated, and therefore the differentiations can be adjusted functionally to the development of the system.

In summary: segmentation requires co-variation in two dimensions; stratification requires stabilization in three dimensions, and consequently a difference between the reflecting instance and the reflected substance; self-organization generalizes the possibility of reflection using a fourth degree of freedom.

**Self-organization**

In terms of the above spatial metaphor of a cube, one may think of a self-organizing system in terms of alternative cylinders in this cube which the system has available as representations of its identity (*Figure Two*). Each cylinder leads to a different expectation for the composition in the next round. The operation of the *complex* system is uncertain in a fourth dimension with reference to its three-dimensional representations. A self-organizing system is expected to select three-dimensional representations which are functional for its further development. In order to maintain identity in a self-organizing system, both the communications and their co-occurrences have to be selected by the system, *and* to be interpreted self-referentially as information about the *current* state of the system.

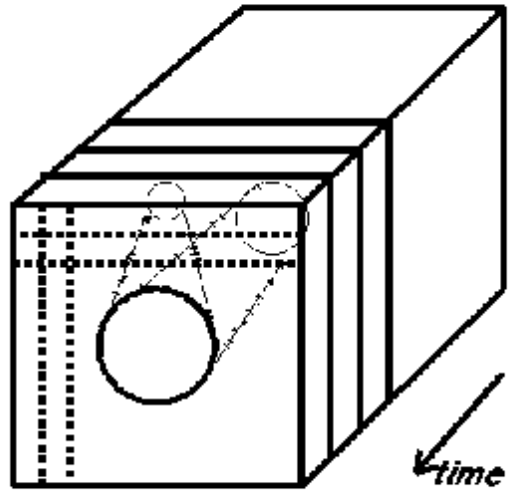

**Figure 2**

*Selection among representations of the past using a fourth degree of freedom*

As noted, observable stabilization can be considered as second-order selection. A next-order selection leads to the potentially global regime of a self-organizing system. 'Variation', 'selection', and 'stabilization' at lower levels can then be considered as sub-dynamics of the complex system, which performs an operation in a 'hyper-cycle', including time and space. Note how the higher-level system *rests* as a selective feedback on top of the lower-level ones; it controls by repressing the possible selections at the lower level.[4]

In the fourth degree of freedom specific resonances between the first-order cybernetics of variation and selection and the second-order cybernetics of variation and stabilization become possible. Simon (1969) has introduced the metaphor of 'locking into resonances' for understanding this evolutionary construction of complex systems. If one of the resonances is entrained through stochastic variation, the functionality of the differentiation spreads, since the remainder of the system consists of groupings that are not yet functional (cf. Smolensky 1986; Kampmann *et al.* 1994). For example, in functionally differentiated organisms, undifferentiated cells cease to occur. Analogously, in the late Middle Ages, the Investiture Controversy between the Pope and the Emperor led to a differentiation among hierarchies in the stratified organization of society. Among other things, this induced the transition from a stratified high-culture to a functionally differentiated society (Luhmann 1989).

**Translations**

If the non-differentiated communication is indicated with *j* (see above), the differentiated medium must be indicated with *jk*, since a grouping variable *k* is added. If the one functionally differentiated subsystem, which for example communicates in *j* and *k1*, communicates with another functionally differentiated subsystem (in *j* and *k2*) of the same system, this does not imply de-differentiation and thus communication in only *j*, but the emergence of communication among *j*, *k1*, and *k2*. De-differentiation among subsystems can occur only *locally*, when *k1* and *k2* cancel one another like in patterns of interference. In general, evolutionary integration means an increase in complexity; only local specimens can be found that are not yet differentiated, and therefore are able to carry the next generation.

For example, if a human being wishes to move, he or she needs an interface which not only makes the organs involved (nerves, muscles, bones, etc.) recognize one another as tissue of the same animal (*j*), but which also structures the communication between, for example, the nervous system (*jk1*) and the motoric apparatus (*jk2*). This operational coupling requires the coupling into an interfacing system (e.g., a synapse) which 'knows' how to translate input into output; by structurally doing so, the interfacing system composes a three-dimensional system. Only a three-dimensional communication system (*jkl*) can contain sufficient complexity to perform translations between differentiated systems.

Analogously, language can be considered as the yet undifferentiated medium of communication in the social system (*j*). Differentiation attaches a suffix *k* to all usages of language. Following this differentiation one is no longer able to compare two communications in terms of a single medium of communication, e.g., a common language, without (initially local) reflection on the contextual meaning of the communication. In a stratified social system, social communication can still be integrated, since only the stable center is eventually allowed to reflect on the meaning of a communication. In this case, the differentiation between the function (*k*) and the meaning (*l*) of a communication (*j*) remains repressed. However, as the social system becomes functionally differentiated in its organization, three-dimensional subsystems of communication (*jkl*) allow the carrying agents to operate in terms of translations of (*a priori*) input into (*a posteriori*) output. A differentiated communication system needs reflexive agency among its subsystems, since it would otherwise fall apart.

A translation is formally equivalent to a reflection: one can 'fix' the system as a communication channel and consider it in terms of relations between input and output (Shannon 1948) or deconstruct the same system as a reflector that uses three degrees of freedom. If the communication channel, however, is no longer fixed, it is expected to change, among other things, its reflexive function (if only by wear and tear). Self-organization is an analytical consequence of replacing the assumption about fixed channels that can transmit with more or less noise, with communication systems that themselves may change when disturbing the transmission. The declaration of the additional context of the communication provides us with a dual perspective: one can focus on input/output relations or consider the input as contextual disturbances of an evolving system that provides its environment with an output by exhibiting change (see *Figure Three*).

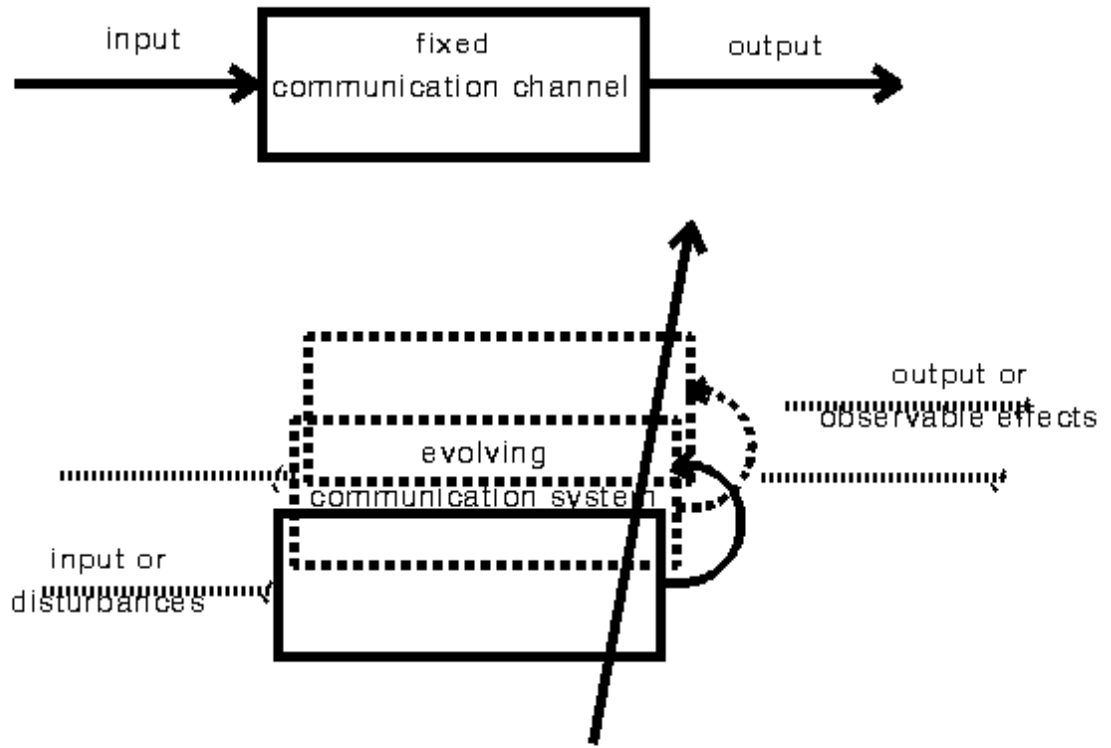

**Figure 3**

*A "fixed" communication channel and an evolving communication system*

**The Duality of Social Communication**

Reflexive systems are able to communicate among themselves, since they can bounce information back and forth if they relate to the same medium of communication (Maturana 1978). Self-organizing systems are in principle competent to communicate in *two* dimensions at the same time, since they have one more degree of freedom for the reconstruction. (One degree of freedom in a translation has to be used by the receiving system for declaring the noise generated by the transmitting system.) Human language is the evolutionary achievement that allows for the communication of information and the meaning of the uncertainty in the same communication.

In other words, information and meaning can be considered as dimensions of human communication. The difference between the information and meaning in human communication has been codified in language: language can hold information, and therefore translation allows us to redirect the information into a next communication, but reflexively. Note that a reflexive communication cannot be unambiguously communicated in the same act as the substantive communication on which it reflects, but only with hindsight. The analytical reason for this duality in social communication is the need to declare one dimension of the communication as noise. Given two channels of a three-dimensional translation (see above), one can either combine the substance of the communication with its function (context) or the substance with its meaning (over time).

Two messages in three dimensions using the same medium contain sufficient redundancy for a four-dimensional human mind in order to reconstruct (i) the expected information content, (ii) the transformation by the media through which it passed (i.e., the contextual information), (iii) the intended meaning of the communication. Obviously, the necessary filtering of the noise requires a memory function that is structurally allowed the freedom to relate the various reflections internally.

Thus, a reflexive memory function at the address of an actor is needed for the social translation, since the social system has to operate at least twice before it unambiguously communicates both substantively and reflexively (cf. Luhmann 1990).[5] But the reflexive actors are distributed, and therefore the various reflections can be communicated. A global system of translations can be organized as a cultural evolution on top of stratified processes of social communication when the reflexive layer of communications is differentiated as a degree of *freedom* at the level of the social system. This presumes a certain complexity (like in bourgeois cities and Protestant churches) so that the different roles in the communication can be distinguished in terms of selections (as opposed to preordained roles).

**The Regime of Modernity**

When the reflection is no longer fixed (like in a stratified society), the translation becomes a historical variable. Thus, the transition to modernity affects the nature of social communication: the unit of social communication, which may contain both information and meaning (*jk*), is extended with the historical context (*l*). Henceforth, communications are expected to translate among systems of translations. No common language tends to be left among the differentiated (sub-)systems to which one may hope to recur for the system's integration (cf. Habermas 1981).

From this perspective, the translation of the Bible into the various national languages may have been a crucial step in the formation of 'modernity'. Historically, the reflexive function, i.e., the attribution of meaning, could be uncoupled from the hierarchical centre of the stratified system ('Rome') because of the emphasis in the Gospel on *personal*, i.e., decentralized, salvation.[6] Whereas in the Middle Ages a personal or local interpretation of the *imitatio Christi* might easily lead to 'ex-communication', God's Word could in the long run provide semantic leverage for breaking the hierarchical fixation of the Roman system in the reflexive dimension. In Protestantism each individual is equal before God; the World is given to people as a latent structure in their network of relations.

It goes beyond the framework of this paper to elaborate on this evolutionarily recent transition from a stratified high-culture into the modern regime that is based on rewarding translations. When the communication in the social system can be *stabilized* in an hierarchy, a high-culture first develops. However, the social system can be developed in the fourth dimension whenever communicating agents, reflexively aware of their different positions in the social system, begin to systematically use the degree of freedom between differentiation and reflexivity as another dimension in their communication. The evolutionary achievement is the freedom to internally adjust the reflections in relation to local exigencies of differentiated developments.

Protestant ethics sanctioned local reflexivity in our colloquial command of the world. If reflexive functions can subsequently be ascribed as degrees of freedom to subsystems of communication (e.g., the free market, the freedom of religion, etc.), the gradual transition into a self-organizing regime of translations is only a matter of time: global adjustments are based on the recursive selection of lower-level variations.

**The Endogenous Character of Technological Change**

A system can deconstruct a signal of one-lower dimensionality than it has available, since it needs the additional dimension in order to provide the signal with a value, and thus to estimate the noise. In the second dimension of the uncertainty we have called this selection the positioning of the relational information; in the third dimension reflection; and in the fourth dimension it has occasionally been called reflexivity, but the operation can be clearly distinguished from reflection by calling it self-organization (see *Table One* above). The underlying operation among the various dimensions, however, is identical: the incoming information is always mutual information with reference to the expected information content of a communication system. The receiving system reconstructs the signal by normalizing the incoming information in its own terms. This presumes an internal representation, and hence the projection onto another dimension of the system's operation (Leydesdorff 1992).

A communication system that operates in three dimensions can reconstruct input into output; a four-dimensional system can reconstruct in three dimensions. For example, a bird can build a nest in three dimensions. It does so instinctively, but not reflexively. Technological artifacts, however, are based on discursive reconstructions, and should thus be attributed to the social communication system. As noted, the reflexive system therefore can carry a cultural evolution on top of the natural one.

Given the above definitions, the system of reference for the cultural evolution of a modern social system, is no longer an individual or a community as an aggregate of individuals, but the reflexive interaction system or discourse *among* individuals. The interaction terms are based on the *grouping* as another dimension than the grouped one (see above). While a stratified system can be organized in terms of aggregated communities, this system can be evaluated as contingent as soon as the interaction terms between different groups account for a larger part of the uncertainty that prevails than the sum of the within-group variances.[7]

The capacity to communicate reflexively about the grouping, and then reflexively to reorganize the communication locally into another (provisional) stabilization emerged during the reformation, but it was institutionalized during the scientific revolution of the 17th century. The experimental condition is almost paradigmatic for the local reconstruction (cf. Latour 1983; Shapin & Schaffer 1985). The reflexive recombination --- i.e. (re)attribution of grouping --- endogenizes technological change in the social system. While pre-modern societies had a set of institutions and techniques available which were specific and provisionally stabilized, technological change is a characteristic of a self-organizing social system. Translations among systems of translation enable the reflexive carriers to explore possible recombinations. As noted, the differentiated system tends to build structurally on those instantiations which are functional for its further development (cf. Simon 1969; Swenson 1989). As functional differentiation is increasingly inscribed into the system, redistributions among asynchronous developments force the system to higher-order innovations, periodically (cf. Schumpeter 1943).

Remember that the social system is not a given (like a biological body), but it remains a construct that can be reconstructed, in principle. Therefore, the evolutionarily advantageous reorganizations can be rapidly recognized. These innovative dynamics are nowadays a sub-cybernetics of the social system; cultural evolution has become sustainable only as far as it is innovative. (Otherwise, the social system would have to 'die' like 'natural' systems.) In a technological culture, however, 'nature' is continuously reconstructed in terms of changing patterns of communication. Furthermore, cultural evolution reinforces itself by producing a continuous stream of technological artifacts, since it is based on recursive optimization. Thus, the transformation of 'nature' has become a functional part of the cultural system. From this perspective, it might make an 'environmental' difference at the global level if the criteria for these optimizations could be made the subject of higher-order theoretical reflections (cf. Freese 1988).

**Differentiation among Reflexive Discourses**

Would we not need a fifth dimension for a meta-theoretical reflection? How can one explain that we are able to understand self-organization, while being self-organizing systems ourselves? One expects that a four-dimensional system is able to reconstruct a three-dimensional system to the extent that it can design one. However, a four-dimensional mind can only reconstruct a four-dimensional system to the extent that it can develop and improve its mental mappings of it, given the specification of a perspective (Hinton *et al.* 1986). Thus, we are not able to construct a dynamic social system like we are able to engineer an electronic circuit, but we are able to reflexively understand the dynamics of the social system by taking a point of reference. As Luhmann (1984) argued, this point of reference for the reflection is to be considered as the analyst's 'blind spot'.

Can one recursively reflect on one's blind spot? Kuhn's (1962) metaphor of paradigms has been particularly helpful in understanding changes in perspectives on mental mappings of complex systems. Paradigm changes enable us to change back- and foreground like in a Gestaltswitch, but at the supra-individual level. Each paradigm allows the participants to communicate reflexively, since a perspective is provisionally stabilized. Problems which cannot be clearly communicated in one paradigm may be solvable after a paradigm switch. Kuhn, however, considered the communication among paradigms as virtually impossible.

Like in a belief system, a paradigm provisionally fixes the fourth degree of freedom in the communication by making one preferential selection among the various possible perspectives on the complex system(s) under study. The various paradigms take other axes for the reflection, and therefore they can be considered as 'incommensurable'. However, the paradigms compete in their efforts to understand the subject of study. Thus, competing theories constitute a layer of reflexive systems of communication on top of the complex dynamic systems under study.

A reflexive analyst is able to understand a paradigm switch, because a four-dimensional mind is able to translate among translation systems, in principle. Thus, the various paradigms remain only '*nearly* incommensurable' in the sense that an evolutionary system remains nearly decomposable (Simon 1973). The more the main axes of the reflection are orthogonal, the more the paradigms are expected to have grown incommensurable. The significantly less frequent interaction terms between these differentiated theoretical systems, however, are expected to organize the functionality of the theoretical discourses for the evolutionary transformation of the systems under study.

**The 'Duality' in the Sociological Understanding**

How can this reasoning help us in clarifying the differences among paradigms in sociology? Let us assume that the social system is indeed a complex and dynamic system. In order to capture as much complexity of the system under study as possible, comprehensive theories are expected to develop increasingly along orthogonal dimensions. In general, a four-dimensional system has four orthogonal projections in three-dimensional spaces. Thus, one should expect three fully fledged sociologies and one meta-theoretical reflection to become dominant metaphors for reconstructing the dynamics of the social system. The three (ideal-typical) sociologies are expected not to take into consideration either (*i*) the structural dimension of differentiation, (*ii*) reflexivity in the time dimension, or (*iii*) the re-attribution of meaning in the fourth dimension. Meta-theoretical discourse abstracts from the substance of these sociologies (in the primary dimension of variation), and postulates a mechanism for their integration.

The three expected sociologies can be identified as (*i*) historical approaches that tend to consider agency as a source of yet undifferentiated variation; (*ii*) systems-theoretical approaches that tend to focus on the invariants of the system, and thereby neglect the dynamics among the dimensions over time (e.g., Parsons' structural-functionalism); and (*iii*) symbolic interactionism that neglects the difference between substantive and reflexive communications by considering all social communications under the single perspective of 'meaning'. In their ideal-typical forms these sociologies exclude one another (cf. Grathoff 1978). For evolutionary reasons, however, one expects *near* decomposability in practice. Consequently, the discursive translations among these three discourses are expected to develop theoretical perspectives with a significantly lower frequency than within each of the paradigmatic discourses (Simon 1973).

In other words, the fourth (meta-theoretical) discourse is evolutionarily 'later'. It has hitherto been developed by using mainly an anthropomorphic metaphor: the reflexive analyst is supposed to be able to integrate the various discourses in a meta-theoretical reflection. However, this formulation reduces the problem to a psychological or a philosophical one (e.g., Woolgar 1988). The sociological problem is again how this reflexivity can be communicated.

Recently, parallel and distributed processing has provided us with the mental model to understand this operation (e.g., Rumelhart *et al.* 1986; Leydesdorff 1993). Each sociology can be considered as an independent processor of a discourse, but the 'intertextuality' is generated by the program that runs in the network among these discourses. The texts are expected to appear with different frequencies in the (hyper-cyclic) intertextuality (cf. Kristeva 1980). Furthermore, the spectra of frequencies are expected to change historically as a result of the interactions. The specific sociologies provide us with theories about how the various loops operate; the study of the intertextuality among the various descriptions should provide us with an expectation about their relative weights.

Is one able to specify the production rule for 'intertextuality' among the three sociologies specified above? Is it possible to achieve a higher-order dimensionality in the understanding, i.e., one that allows for a duality (Giddens) or perhaps a higher-order plurality in the discursive representation of a four-dimensional system without generating confusion? The crucial point is that one would need a representation which is dynamic in terms of choosing a perspective (as in a movie). By using a spatial depiction or a geometrical metaphor in the narrative one is not able to represent change in the data and in the relevant categories for organizing the data in the same pass. In general, the scholarly discourse tends to become confused without a clear distinction between *a = non-a* as a categorical contradiction and a permitted change in the value of a variable.

Algebra enables us to distinguish dynamically between a change in a variable and its expected value, since the variable ($x$) can be redefined as a flux ($dx/dt$). More specifically, the algorithmic simulation provides us with the dynamic representation (on the screen) which visualizes change both in structure and in the data like in a movie, while at the same time we can keep track of the cause of the observable effects in terms of the underlying computer code. In other words, the recursive algorithm allows us to distinguish between the observable dynamics of the macro-system and the micro-variation at lower levels in nested layers of selective conditions without becoming confused. The behaviour of the model system is *more* complex than the composing sub-dynamics, while only the latter can be made the subject of substantive theorizing (Langton 1989). The computer code can be considered as the genotypical specification of the phenotypical behaviour of a model on the screen; the resulting model captures the interaction terms 'in between' the discursive representations.

Without theoretical assumptions the problems are non-computable, since each additional context introduces an infinite number of possible interactions. Substantive specification operates as a selection device among other selection devices; the various specifications condition one another in the model system. As noted, the social system is not a biological given but a reflexive reconstruction, and thus social communications allow for complexities in the communication with a complexity of even higher than four. For example, one can consider three or more communications as a single operation of the social system, analogously to the above specified possibility to operate in iterations of two communications. Cultural evolution can develop increasingly complex patterns to the extent that (some) actors are able to carry the reconstructive communications.

In summary, common languages allow for one layer of reflexivity without confusion. Codification in specialist languages allows for the (provisional) stabilization of meaning in nearly decomposable reflexive layers of communication (e.g., 'paradigms'). Codified computer code allows for higher-order recursion. Of course, an interpreter may need a language for the theoretical appreciation of the simulation results, but this should not obscure the evolutionary constraints on 'human' understanding in 'natural' languages.[8]

**Implications for Sociological Theorizing**

As noted, a reflexively specified translation requires at least two communications. By accepting methodological assumptions about the organization of the inference, reflexive discourses therefore can become codified. Given the implied blind spot, all inferences thus generated remain necessarily uncertain with respect to the relative relevance of the perspective for understanding the system(s) under study.

Hitherto, the issue of the selective perspective has been elaborated in the social sciences primarily as a methodological and self-referential concern about bias. For the algebraic transformation of theorizing, however, it should be discussed as a choice of a relative perspective in relation to a space of other possible perspectives. First-order theoretical results can then be formulated as conditional statements ('selection') about probability distributions ('variation') that can be specified in algorithmic code as sub-routines. By relating the mechanisms and translations in specific orders, one is able to specify the consequences of their interactions in terms of ranges of expectations (cf. Hanneman 1988).

Substantive theories provide us with the specification of an uncertainty that cannot be provided by the mathematics. One needs discursive theorizing to specify the nature of the mechanisms of variation and selection. These mechanisms can be considered as the building blocks of the higher-order cybernetics. In general, the specification of the sub-dynamics generates a multi-dimensional problem space. Each theoretical reconstruction is partial, and must be positioned with reference to this problem space. The mathematical model, however, can be studied without a relation to specific instantiations in the system under study. Therefore, it allows one to generalize from the specifications to the problem space, and in principle to suggest states other than those which are intuitively accessible. Thus, the system of theoretical representations is able to bootstrap from specification to generalization.

In my opinion, the possibility of understanding *the dynamics among theories* as another layer of complex (reflexive) systems on top of the complex systems under study, has recently emerged on the basis of the evolution of those sciences that can be reconstructed in terms of computer code (e.g., Langley *et al.* 1987; Hanneman 1988; Freese 1988; Anderson *et al.* 1988; Langton 1989; Andersen 1992 and 1994; Leydesdorff 1995b; Langton *et al.* 1992). As far as these meta-theoretical reconstructions can again be communicated, they indicate the possibility of a general theory of communication.

Let us first focus on the sociological relevance of this understanding, and postpone the discussion of the epistemological status of a general theory of communication to the next section. By scanning the range of possibilities using a computer model, the 'unintended consequences' of specifications can be made visible in terms of expectations. This option generalizes Giddens' (1979 and 1984) notion of 'unintended consequences'. The 'unintended consequences' are phenotypical outcomes, while the specifications remain genotypical. Any observable system under study will systematically generate 'unintended consequences' if looked upon reflexively, since a system 'in the making' contains an emerging dimension.

One is now able to generalize this insight with reference to theorizing: the dual perspective of the participant and the analyst or --- in other words --- the relations between substantive specification and formal modelling generate a tension between language and meta-language that drives the scientific research process. More specifically, the theoretical specifications are discursive reflections on the resonances that can be observed historically. The algebraic understanding in terms of formal models and fluxes can be developed into a discourse of a different order than the substantive theories that went into its construction. The hyper-cyclic model then may guide us in the search for alternative development patterns, since it recursively recombines these reflexive insights in other possible orders.

In summary, by specifying theoretical expectations about the relevant dimensions and the how of the interactions, the analyst is able to reduce the computational complexity. Special theories make new problems computable, in principle. The substantive elaboration challenges the further development of the mathematics involved, since it limits the number of relevant interaction terms. Simulations can help us to define more precisely the complexity in terms of interacting procedures; simulation results challenge the appreciative understanding of the dynamics at a subsequent stage.

**Conclusions**

The recursivity of the selection in communication has been crucial to the argument. This assumption has enabled us to clarify long-standing problems like the difference between differentiation and stratification, information and meaning, or reflection and self-organization in terms of a single principle. (See *Table One* for a summary of the various concepts.) The non-linear dynamics of self-organizing systems are applicable to (*i*) social systems, (*ii*) to how these systems can reflexively be understood in terms of sociologies, and (*iii*) to the meta-theoretical understanding of the dynamics among the relevant sociologies.

The concept of reflection in terms of hierarchical layers should be replaced with 'reflection' as an orthogonal dimension of the complex construct. The operationalization of reflection as a recursion of the selection allows us epistemologically to formalize 'reflexivity' without ontologically reifying it as a substantively higher-level. Thus, this approach enables us thoroughly to solve the so-called reflexivity problem in post-modern sociology, i.e. the problem that one cannot claim priority for a specific reflection concerning reflexive actions. *If reflection is a contingent property of the communication, it is possible to ask for the quality of a reflection*. The formalization, however, remains in need of substantive specification, for example, in terms of reflexive discourses.

What kind of theory of communication might the implied meta-sociological understanding provide? One would expect a general theory of communication to encompass the mathematical theory of communication and the special theories of communication at the hyper-cyclic level. However, would one be able to understand this theory in substantive terms or only as a formal possibility? In my opinion, the recursivity of the operation provides us with a metaphor for the understanding.

Remember that social structure is not a given, but an expectation on the basis of a theoretical assumption. Accordingly, the operationalization of the social interaction can change with the perspective or with the further development of sociological theory. Analogously, sociological discourses can be modified reflexively by a next-order interaction. In other words, the simulations inform us about 'unintended consequences' of earlier specifications, and thus, they can help us to recursively improve these specifications (like in the case of an update).

The simulation results provide us with a representation of the super-system. On the one hand, the algorithmic specification in terms of sub-routines allows us to backtrack from the simulated phenomena to the theoretical specifications. On the other hand, the model reconstructs the theoretical understanding. The operation is so recursive that the analyst would have to sacrifice explanatory power in order to stabilize the theoretical appreciation of the results in the same discourses as the *a priori* expectations. From this perspective, a theoretical explanation has the status of a translation of the hypotheses.

---

**Notes**

1. Some authors (e.g., Brillouin 1962; Bailey 1990) have defined this difference ("negentropy") as information. When the focus is no longer on a fixed communication channel, but on an evolving communication system, one should distinguish between the expected information content of the receiving system, and *observed* information that is positioned by this system in a subsequent update. The observing system can meaningfully position the incoming information with reference to its *previous* state. It can be shown that with reference to the *a priori* system the probabilistic entropy of the interaction may sometimes have a negative value, and therefore add to the *redundancy* of the receiving system (cf. Leydesdorff 1992 and 1995a). With reference to the *a posteriori* situation, however, the uncertainty has always to increase (Georgescu-Roegen 1971: 410ff.; cf. Hayles 1990).

2. The number of possible states of a network system increases with the number of nodes in the exponent. Thus, the number of possible states is non-polynomial complete; it becomes rapidly uncomputable with an increasing number of nodes.

3. If all possible links would be used to the same extent and at the same time, the network would be completely uncertain. As an entropical system, the network would then have "died".

4. For example, if a muscle is denervated, control by the higher-level system is released, and the sensitivity of the lower-level system for disturbances is increased.

5. A communication system among animals would have to operate at least three times before a community (e.g. a population of insects) can be generated. The assumption of higher-order hyper-cycles is more common in modelling biological systems (cf. Langton et al. 1992). However, the insects themselves are not expected to carry higher-order memory functions.

6. The duality of the communication is paradigmatically entailed in the intertextuality between the Old and New Testament. The New Testament reflected the Personal meaning of the substantive Communication in the Old one. Christianity, however, initially adapted to the format of the Roman system, which spanned the whole world (kat' holen gen or Catholic), but in terms of a stable Empire.

7. Although differently defined, 'variance' and 'information content' are both measures of the uncertainty, and therefore semantically equivalent (Theil 1972).

8. One expects that a five-dimensional system is able to reconstruct and to design a four-dimensional one. Thus, the project of artificial evolution is tractable, but the problem is currently uncomputable given the availability of hard- and software (cf. Leydesdorff 1994).

**References**

Andersen, E. S. 1992. *Artifical Economic Evolution and Schumpeter*. Aalborg: Institute for Production, University of Aalborg.

Andersen, E. S. 1994. *Evolutionary Economics. Post-Schumpeterian Contributions.* London: Pinter.

Anderson, P. W., K. J. Arrow, and D. Pines (eds.) 1988. *The Economy as a Complex Evolving System*. Reading, MA: Addison-Wesley.

Arthur, W. B. 1988. 'Competing Technologies'. Pp. 590--607 in *Technical Change and Economic Theory*, edited by G. Dosi, C. Freeman, R. Nelson, G. Silverberg, and L. Soete. London: Pinter.

Bailey, K. D. 1990. 'Why *H* does not measure information: the role of the "special case" legerdemain solution in the maintenance of anomalies in normal science'. *Quality and Quantity* 24: 159--71.

Brillouin, L. 1962. *Science and Information Theory*. New York: Academic Press.

Burt, R. S. 1982. *Toward a Structural Theory of Action*. New York: Academic Press.

Dosi, G. 1982. 'Technological paradigms and technological trajectories'. *Research Policy* 11: 147--62.

Freese, L. 1988. 'Evolution and Sociogenesis', in: E. J. Lawler and B. Markovsky (eds.), *Advances in Group Processes*, Vol. 5. Greenwich, CT: JAI Press, pp. 53--118.

Georgescu--Roegen, N. 1971. *The Entropy Law and the Economic Process*. Cambridge, Mass.: Harvard University Press.

Giddens, A. 1979. *Central Problems in Social Theory*. London: Macmillan.

Giddens, A. 1984. *The Constitution of Society.* Cambridge: Polity Press.

Grathoff, R. (ed.) 1978. *The Theory of Social Action. The Correspondence of Alfred Schutz and Talcott Parsons*. Bloomington and London: Indiana University Press.

Habermas, J. 1981. *Theorie des kommunikativen Handelns*. Frankfurt a.M.: Suhrkamp.

Hanneman, R. 1988. *ComputerAssisted Theory Building: Modeling Dynamic Social Systems*. Beverly Hills: Sage.

Haraway, D. 1988. 'Situated Knowledges: The Science Question in Feminism and the Privilege of Partial Perspective', *Feminist Studies* 14: 575--99.

Hayles, N. K. 1990. *Chaos Bound; Orderly Disorder in Contemporary Literature and Science*. Ithaca: Cornell University Press.

Hinton, G. E., J. L. McClelland, and D. E. Rumelhart, 1986. 'Distributed Representations'. Pp. 77--109 in (Rumelhart *et al.* 1986).

Kampmann, C., C. Haxholdt, E. Mosekilde, and J. D. Sterman, 1994. 'Entrainment in a Disaggregated Economic Long-Wave Model'. Pp. 109--24 in: L. Leydesdorff and P. Van den Besselaar (eds.), *Evolutionary Economics and Chaos Theory: New Directions in Technology Studies*. London: Pinter.

Kaufer, D. S., and K. M. Carley, 1993. *Communication at a Distance*. Hove/London: Erlbaum.

Krippendorff, K. 1994. 'A Recursive Theory of Communication'. Pp. 78--104 in: D. Crowley and D. Mitchell (eds.), *Communication Theory Today*. Cambridge, UK: Polity Press.

Kristeva, J. 1980. *Desire in Language*. New York: Columbia University Press.

Kuhn, T. S. 1962. *The Structure of Scientific Revolutions*. Chicago: University of Chicago Press.

Langley, P., H. A. Simon, G. L. Bradshaw, and J. M. Zytkow 1987. *Scientific Discovery. Computational Explorations of the Creative Processes*. Cambridge, Mass./ London: MIT Press.

Langton, C. G. (ed.) 1989. *Artificial Life*. Redwood City, CA: Addison Wesley.

Langton, C. G., C. Taylor, J. D. Farmer, and S. Rasmussen (eds.) 1992. *Artificial Life II*. Redwood City, CA: Addison Wesley.

Latour, B. 1983. 'Give Me A Laboratory and I will Raise the World', in: K. D. Knorr-Cetina and M. J. Mulkay (eds.), *Science Observed*. London: Sage, pp. 141--70.

Lazarsfeld, P. F., and N. W. Henry 1968. *Latent structure analysis*. New York: Houghton Mifflin.

Leydesdorff, L. 1992. 'Knowledge Representations, Bayesian Inferences, and Empirical Science Studies'. *Social Science Information* 31: 213--37.

Leydesdorff, L. 1993. '"Structure"/"Action" Contingencies and the Model of Parallel Distributed Processing'. *Journal for the Theory of Social Behaviour* 23: 47--77.

Leydesdorff, L. 1994. 'The Evolution of Communication Systems', *Systems Research and Information Science* 6: 219--30.

Leydesdorff, L. 1995a. 'The Production of Probabilistic Entropy in Structure/Action Contingency Relations', *Journal of Social and Evolutionary Systems* 18: 339--56.

Leydesdorff, L. 1995b. *The Challenge of Scientometrics: The development, measurement, and self-organization of scientific communications*. Leiden: DSWO Press, Leiden University.

Luhmann, N. 1984. *Soziale Systeme. Grundriß einer allgemeinen Theorie*. Frankfurt a. M.: Suhrkamp.

Luhmann, N. 1989. *Gesellschaftsstruktur und Semantik. Vol. III*. Frankfurt a.M.: Suhrkamp.

Luhmann, N. 1990. *Die Wissenschaft der Gesellschaft*. Frankfurt a.M.: Suhrkamp.

MacKay, D. 1969. *Information, Mechanism, and Meaning*. Cambridge: MIT Press.

Maturana, H. R. 1978. 'Biology of Language: The Epistemology of Reality'. Pp. 27--63 in *Psychology and Biology of Language and Thought. Essays in Honor of Eric Lenneberg*, edited by G. A. Miller and E. Lenneberg. New York: Academic Press.

Rumelhart, D. E., J. L. McClelland, and the PDP Group 1986. *Parallel Distributed Processing*, Vol. I. Cambridge, MA/London: MIT Press.

Schumpeter, J. 1943. *Socialism, Capitalism and Democracy*. London: Allen & Unwin).

Shannon, C. E. 1948. 'A Mathematical Theory of Communication'. *Bell System Technical Journal* 27: 379--423 and 623--56.

Shapin, S., and S. Schaffer, 1985. *Leviathan and the Air-pump: Hobbes, Boyle, and the Experimental Life*. Princeton, N.J.: Princeton University Press.

Shinn, T. 1987. 'Géometrie et langage: la structure des modèles en sciences sociales et en sciences physiques'. *Bulletin de Méthodologie Sociologique* 1 (nr 16): 5--38.

Simon, H. A. 1969. *The Sciences of the Artificial*. Cambridge, MA/ London: MIT Press.

Simon, H. A. 1973. 'The Organization of Complex Systems'. Pp. 1--27 in *Hierarchy Theory. The Challenge of Complex Systems*, edited by H. H. Pattee. New York: George Braziller.

Smolensky, P. 1986. 'Information Processing in Dynamical Systems: Foundation of Harmony Theory'. Pp. 194--281 in (Rumelhart *et al.* 1986).

Swenson, R. 1989. 'Emergent Attractors and the Law of Maximum Entropy Production: Foundations to a Theory of General Evolution'. *Systems Research* 6: 187--97.

Theil, H. 1972. *Statistical Decomposition Analysis*. Amsterdam: North-Holland.

Woolgar, S. 1988. *Science. The very idea*. London/New York: Tavistock Publications.